# Factored Filtering of Continuous-Time Systems

**E. Busra Celikkaya**
University of California, Riverside
celikkae@cs.ucr.edu

**Christian R. Shelton**
University of California, Riverside
cshelton@cs.ucr.edu

**William Lam**
University of California, Irvine
willmlam@ics.uci.edu

## Abstract

We consider filtering for a continuous-time, or asynchronous, stochastic system where the full distribution over states is too large to be stored or calculated. We assume that the rate matrix of the system can be compactly represented and that the belief distribution is to be approximated as a product of marginals. The essential computation is the matrix exponential. We look at two different methods for its computation: ODE integration and uniformization of the Taylor expansion. For both we consider approximations in which only a factored belief state is maintained. For factored uniformization we demonstrate that the KL-divergence of the filtering is bounded. Our experimental results confirm our factored uniformization performs better than previously suggested uniformization methods and the mean field algorithm.

## 1 Continuous-Time Markov Systems

We are interested in monitoring (alternatively tracking or filtering) a continuous-time finite-state homogeneous Markovian stochastic system. This implies that evidence and events can arrive at any real-valued time (asynchronously) and be either instantaneous or have real-valued time durations. The system is stationary and has discrete states which means that a sample (trajectory) for the system consists of a series of times at which the system jumped from one state to another.

Such models are common in the queueing theory and verification literatures. The algorithms developed there almost exclusively focus on steady-state properties of the system. A singular exception is the work of Sutton and Jordan (2008) which applied Gibbs sampling to queueing models. By contrast, the continuous time Bayesian network (Nodelman et al., 2002) literature has focused on finite-time properties, so we compare against those algorithms.

In verification, continuous-time models can be used to compute the probability a system will function by a given time, such as in continuous stochastic logic (Baier et al., 2003). Other problems of interest include state estimation in asynchronous systems, such as distributed computer systems (Xu and Shelton, 2010), robotics (Ng et al., 2005), social networks (Fan and Shelton, 2009), or phylogenetic trees (Cohn et al., 2009).

### 1.1 Parameterization

Such a Markovian system is described by an initial distribution over the state space and an intensity (or rate) matrix, often denoted $Q$. The diagonal element $q_{ii} = -q_i$ where $q_i \geq 0$ is the rate of leaving state $i$. This means that the density of the duration of the process remaining in state $i$ for exactly a duration $\Delta t$ is an exponential distribution: $q_i e^{q_i \Delta t}$. The non-diagonal elements, $q_{ij} \geq 0$ are the rates of transitioning from state $i$ to state $j$. $q_i = \sum_j q_{ij}$ in a closed system. (The rate of leaving a state is equal to the sum of the rates of moving to any other state.) The probability of transitioning to state $j$ immediately upon leaving state $i$ is $q_{ij}/q_i$. Note that the diagonal elements of $Q$ are non-positive and the non-diagonal elements are non-negative. The sum of each row is 0.

Monitoring such a system consists of keeping track of the probability distribution over the state at the current time $t$, given all evidence prior to $t$. We would like a recursive solution in which evidence before $t$ can be discarded once the distribution (or its estimate) at $t$ has been computed. As an initial problem, we will be interested in tracking in the absence of any evidence. We will cover how to incorporate evidence in Section 5.

To simplify notation, we will assume that we have the distribution at time 0 and wish to propagate this dis-tribution to time $t$. If $p$ is the distribution at time 0, represented as a row vector, then $p'$, the distribution at $t$ also represented as a row vector, is

$$p' = pe^{Qt}$$

where $e^{Qt}$ is the matrix exponential of $Qt$. For a few hundred states, this computation is tractable. Yet, the number of states of a system grows exponentially with the number of properties required to describe the system's state. Very quickly it becomes impossible to represent the matrix $Q$ or even the result $p'$ exactly. It should be noted that almost any useful structure that might be imposed on $Q$ is destroyed by the matrix exponential. Kronecker-sum constructions of $Q$ are the sole exception, but they represent processes with completely independent sub-processes and are therefore not interesting on their own.

Previous results have looked at sparse representations of $p$ and $p'$ (Sidje et al., 2007). These work well for systems with very tightly coupled components in which there are only a few joint assignments to the system properties that have large probability at any one time. By contrast, in this work we are interested in systems with many (relatively) weakly interacting components. Therefore, instead of approximating $p$ and $p'$ with sparse vectors, we will approximate them with factored representations.

In order to have a factored representation, any state must be able to be described as a set of properties or variables. We let $X_i$ represent variable $i$ (out of $n$); an assignment to $X_i$ is denoted $x_i$; and therefore a joint assignment (a complete state) $\boldsymbol{x} = (x_1, \ldots, x_n)$. We will, therefore, be considering the case of calculating $p' = pe^{Qt}$ when $p$ is approximated as $\hat{p}(\boldsymbol{x}) = \prod_i \hat{p}_i(x_i)$ and $p'$ is similarly approximated as $\hat{p}'(\boldsymbol{x}) = \prod_i \hat{p}'_i(x_i)$.

**Example.** *Our running example will use a 4-state system described by the rate matrix*

$$Q = \begin{bmatrix} -4 & 1 & 3 & 0 \\ 2 & -7 & 0 & 5 \\ 4 & 0 & -5 & 1 \\ 0 & 6 & 2 & -8 \end{bmatrix} .$$

*If the distribution at $t = 0$ is $p = \begin{bmatrix} 0.4 & 0.1 & 0.2 & 0.3 \end{bmatrix}$, the marginal distribution at $t = 0.5$ is*

$$p' = pe^{0.5Q} = p \begin{bmatrix} 0.4313 & 0.1443 & 0.3098 & 0.1147 \\ 0.2958 & 0.2648 & 0.2221 & 0.2173 \\ 0.4106 & 0.1406 & 0.3305 & 0.1183 \\ 0.2886 & 0.2622 & 0.2293 & 0.2199 \end{bmatrix}$$

$$= [0.3708 \quad 0.1910 \quad 0.2810 \quad 0.1572] .$$

## 2 Matrix Exponential Calculations

Moler and Loan (2003) give an excellent description of the numeric difficulties in calculating the matrix exponential. We concentrate here on two of the principal methods for calculating $pe^{Qt}$.

### 2.1 ODE

One method is to rewrite $f(t) = pe^{Qt}$ as the ordinary differential equation

$$\dot{f}(t) = f(t)Q \qquad f(0) = p$$

and solve using ODE integration methods. The most common ODE solver is the Runge-Kutta-Fehlberg (RKF) method which which adapts the step size to the current error, thereby allowing for quick progress at times of slow system change. The fundamental calculation is the multiplication of a current estimate of the distribution by $Q$ to calculate the time derivative.

### 2.2 Uniformization

Uniformization is a transformation of a continuous-time Markovian system into a discrete-time one. However, it does not correspond to constructing either the embedded Markov chain of the continuous-time process, nor to time-slicing the system at regular intervals. It is equivalent to sampling the intervals between *potential* state changes from an exponential with rate $\alpha$ and then sampling a suitable Markov chain just at these time points (with stochastic matrix $M$), so that the resulting distribution over trajectories matches the original continuous-time Markov system.

Mathematically we can construct $\alpha$ and $M$ as follows. We express the rate matrix $Q$ as

$$Q = \alpha(M - I)$$

so that—provided $\alpha \geq \max_i q_i$—$M$ is a stochastic matrix: all elements are on $[0, 1]$ and the rows sum to 1. Ideally $\alpha$ should be as small as possible as it represents the rate of the process sampling the intervals. Note that while the continuous-time process never has a "self-transition" explicitly, $M$ does have non-zero diagonal elements (corresponding to those states whose rates of leaving are not maximal).

This can be applied to the matrix exponential calculation as

$$e^{Qt} = e^{\alpha(M-I)t} = e^{\alpha tM}e^{-\alpha t} .$$

In the context of the matrix exponential, this transformation is usually performed to remove the negative elements in the resulting Taylor expansion and thereby stabilize the calculation. We wish to calculate

$$pe^{Qt} = pe^{-\alpha t}e^{\alpha tM} = e^{-\alpha t} \sum_{k=0}^{\infty} \frac{(\alpha t)^k}{k!} pM^k . \quad (1)$$

Given that $M$ is a stochastic matrix, $p'$ is a infinite mixture of distributions $p, pM, pM^2, \ldots$ with weights $e^{-\alpha t}, e^{-\alpha t}\alpha t, e^{-\alpha t}\frac{\alpha t}{2}, \ldots$. Each distribution corresponds to $k$ steps of the Markov chain of $M$ and the weights are the probabilities that exactly $k$ steps happen in a time period of $t$. These distributions are calculated recursively, so the important subcalculation is the multiplication of a state distribution by $M$.

**Example.** *Uniformization on the Q matrix results in $\alpha = 8$ and*

$$M = \begin{bmatrix} 0.5 & 0.125 & 0.375 & 0 \\ 0.25 & 0.125 & 0 & 0.625 \\ 0.5 & 0 & 0.375 & 0.125 \\ 0 & 0.75 & 0.25 & 0 \end{bmatrix}.$$

### 2.3 Approximate Versions

Both of these methods involve the calculation of state distributions and their multiplication by either $Q$ or $M$. We will use $v$ to denote any state distribution vector generated during the course of such a calculation. For the ODE method, $v$ is a particular point and $vQ$ is its time derivative. For uniformization, $v$ is an element of the sum and $vM$ is the next element.

While we might assume that the problem specification is compact and thereby assume that $p$ and $Q$ have compact representations, for a system with many variables, it is usually not possible to express intermediate $v$ values exactly, as structure that may exist in $p$ and $Q$ does not exist in $v$. Therefore, the most direct method for constructing an approximate filtering algorithm is to keep $v$ restricted to a smaller representation.

In the case for uniformization, Sidje et al. (2007) proposed keeping a sparse representation for $v$ (by dropping elements less than a given threshold). If $Q$ has a sparse representation (only a small fraction of any row are non-zero) and any given row can be generated as needed, this results in an algorithm with non-exponential running time (in $n$, the number of variables). We call this method sparse uniformization.

Sparse uniformization works well for problems in which the distribution to be tracked is highly peaked. However, in problems in which the variables may be more loosely coupled (and therefore an assignment to one does not necessarily dictate a joint assignment to all), this approximation will either be poor, or it will require a large number of states to be tracked, thereby defeating the quick runtime.

For such systems, a factored representation of $v$ (as suggested above) is more suitable. This naturally suggests two approximate algorithms. First, we might force the RKF integrator to project $v$ to the space of factored distributions. Second, we might force the uniformization method to project $v$ to the space of factored distributions. We call these factored RKF and factored uniformization.

### 2.4 Notation

We will concentrate on approximate calculations of $vM$ ($vQ$ is similar). We consider one subcalculation in the form $v' = vM$. Let $\hat{v}$ be the current approximation of $v$ in factored form: $\hat{v}(\boldsymbol{x}) = \prod_i \hat{v}_i(x_i)$. Similarly, let $\hat{v}' = \hat{v}M$, the result of multiplying $\hat{v}$ by $M$, which is not necessarily completely factored. Finally let $\tilde{v}$ be the projection of $\hat{v}'$ to the set of factored distributions.

Let $M_\perp$ be the operator that both multiplies by $M$ and projects to the space of factored distributions. Thus $\tilde{v} = \hat{v}M_\perp$ and is a factored distribution ($\hat{v}'$ is not).

We abuse notation and let any vector also stand for the distribution it embodies. Further, for any vector $v$, we let $v_i$ be the marginal distribution of $v$ over the variable $x_i$ (even if $v$ is not factored), $v_{-i}$ be the marginal distribution over all variables except $x_i$, $v_{u_i}$ be the marginal distribution over all of $x_i$'s parents, and $v_{i|-i}$ be the distribution over $x_i$ conditioned on all the other variables.

## 3 Factored Rate Matrix

For either factored approximate method, we need to be able to calculate the projection of $vM$ (or $vQ$) onto the space of factored distributions without explicitly generating $vM$ as an exponentially large vector.

While our methods are applicable to Petri nets (Petri, 1962), edge-valued decision diagrams (Wan et al., 2011), and other compact representation of the rate matrix, we focus on using a continuous time Bayesian network (CTBN) to represent $Q$. We note that a correspondence between CTBNs and Petri nets has already been established (Raiteri and Portinale, 2009).

### 3.1 Continuous Time Bayesian Network

A continuous time Bayesian network (CTBN) (Nodelman et al., 2002) has an initial distribution described by a Bayesian network that is not of direct importance to this work. It has a factored representation of the matrix $Q$. Each variable's dynamics depend only on a subset of the other variables (which we denote as parents). Let $\boldsymbol{U}_i$ be the parents of $X_i$. Then for every assignment $\boldsymbol{u}_i$ to $\boldsymbol{U}_i$ there exists an intensity matrix $Q_{X_i|\boldsymbol{u}_i}$ of dimension equal to the number of states of $X_i$. It describes the rates of change for $X_i$ when its parents equal $\boldsymbol{u}_i$. If we let $\delta(\boldsymbol{x}, \boldsymbol{x}')$ be the set of variable indexes for which the assignments in $\boldsymbol{x}$ and $\boldsymbol{x}'$

differ, the complete $Q$ matrix is

$$Q(\boldsymbol{x},\boldsymbol{x}') = \begin{cases} \sum_k Q_{X_k|\boldsymbol{u}_k}(x_k, x'_k) & \text{if } \boldsymbol{x} = \boldsymbol{x}' \\ Q_{X_j|\boldsymbol{u}_j}(x_j, x'_j) & \text{if } \delta(\boldsymbol{x},\boldsymbol{x}') = \{j\} \\ 0 & \text{otherwise.} \end{cases}$$

So $Q$ is sparse (mostly zeros) with non-zero elements on the diagonal and where only one variable changes. It has the form of a sum of factors (compared with a Bayesian network's product of factors).

**Example.** *The $Q$ matrix of our example can be represented by a CTBN of two variables: $A$ (with values $a_0$ and $a_1$) and $B$ (with values $b_0$ and $b_1$). $A$ has no parents and is the parent of $B$. If we let*

$$Q_A = \begin{bmatrix} -1 & 1 \\ 2 & -2 \end{bmatrix}, Q_{B|a_0} = \begin{bmatrix} -3 & 3 \\ 4 & -4 \end{bmatrix}, Q_{B|a_1} = \begin{bmatrix} -5 & 5 \\ 6 & -6 \end{bmatrix}$$

*then the full system has the previous $Q$ matrix if the global states are ordered $a_0b_0, a_1b_0, a_0b_1, a_1b_1$.*

To construct the uniformized matrix $M$, we would like to select $\alpha$ to be the maximally negative diagonal element. In practice this could be a difficult optimization problem. So, instead we select $\alpha = \sum_{i=1}^n \alpha_i$ where $\alpha_i = -\min_{\boldsymbol{u}_i} \min_{x_i} Q_{X_i|\boldsymbol{u}_i}(x_i, x_i)$, the maximally negative diagonal element for $X_i$ for any $\boldsymbol{u}_i$. We let $M_{X_i|\boldsymbol{u}_i} = Q_{X_i|\boldsymbol{u}_i}/\alpha_i + I$ be the stochastic matrix we obtain for $X_i$ with parent assignment $\boldsymbol{u}_i$ using uniformization constant $\alpha_i$. The stochastic matrix $M$ can then be described as

$$M(\boldsymbol{x},\boldsymbol{x}') = \begin{cases} \sum_k \tilde{M}_{X_k|\boldsymbol{u}_k}(x_k, x'_k) & \text{if } \boldsymbol{x} = \boldsymbol{x}' \\ \tilde{M}_{X_j|\boldsymbol{u}_j}(x_j, x'_j) & \text{if } \delta(\boldsymbol{x},\boldsymbol{x}') = \{j\} \\ 0 & \text{otherwise.} \end{cases}$$

where $\tilde{M}_{X_i|\boldsymbol{u}_i} = \frac{\alpha_i}{\alpha} M_{X_i\boldsymbol{u}_i} + \frac{\alpha - \alpha_i}{\alpha} I$. The stochastic matrix $M$ cannot be described as a dynamic Bayesian network (DBN): it has an additive structure, not a multiplicative one, and does not allow the transition of multiple variables. However, $M$ can be described as a mixture of particularly simple DBN transition models: $M = \sum_i \frac{\alpha_i}{\alpha} M_i$ where

$$M_i(\boldsymbol{x},\boldsymbol{x}') = \begin{cases} M_{X_i|\boldsymbol{u}_i}(x_i, x'_i) & \text{if } \boldsymbol{x} = \boldsymbol{x}' \\ M_{X_i|\boldsymbol{u}_i}(x_i, x'_i) & \text{if } \delta(\boldsymbol{x},\boldsymbol{x}') = \{i\} \\ 0 & \text{otherwise.} \end{cases}$$

In particular, it describes a mixture in which with probability $\frac{\alpha_i}{\alpha}$ variable $i$ transitions according to $M_{X_i|\boldsymbol{U}_i}$ and all other variables remain the same. The $i$th mixture DBN has a structure in which all variables $x'_j$, where $j \neq i$, have only $x_j$ (the same node in the previous slice) as a parent. Variable $x'_i$ has the same parents as in the CTBN, with the addition of $x_i$. No intra-slice arcs exist.

**Example.** $\alpha_A = 2$, $\alpha_B = 6$ and so $\alpha = 8$ (as before). We can decompose the same $M$ as before into

$$M_A = \begin{bmatrix} 0.5 & 0.5 \\ 1 & 0 \end{bmatrix}, M_{B|a_0} = \begin{bmatrix} 0.5 & 0.5 \\ 0.67 & 0.33 \end{bmatrix}, M_{B|a_1} = \begin{bmatrix} 0.17 & 0.83 \\ 1 & 0 \end{bmatrix}$$

*which can be viewed as a Markov chain in which with probability $\frac{2}{8}$ $A$ transitions according to $M_A$, otherwise $B$ transitions according to $M_{B|a_0}$ or $M_{B|a_1}$.*

### 3.2 Calculating Factored $vM$

Given the mixture-of-DBN interpretation above of $M$, it is not surprising that $\tilde{v} = \hat{v}M_\perp$ can be calculated in one step without constructing $\hat{v}M$. As $Q$ has the same structure, the same method works for it too.

As projection onto a particular variable, $x_j$, is linear, if $M_{\perp j}$ is the composition of $M$ and the projection onto $x_j$, then $M_{\perp j} = \sum_i \frac{\alpha_i}{\alpha} M_{i\perp j}$ where $M_{i\perp j}$ is the composition of $M_i$ and projection operation onto $x_j$. Furthermore, in $M_i$ only variable $i$ can change:

$$\tilde{v}_j = \hat{v} M_{\perp j} = \sum_i \frac{\alpha_i}{\alpha} \hat{v} M_{i\perp j} = (1 - \frac{\alpha_j}{\alpha})\hat{v}_j + \frac{\alpha_j}{\alpha} \hat{v} M_{j\perp j}.$$

$M_{j\perp j}$ is the marginal over $x_j$ after a transition according to $M_j$. No other variable changes, so this is the expectation of $M_{X_j|\boldsymbol{U}_j}$ over $\hat{v}$'s distribution over $\boldsymbol{U}_j$:

$$\tilde{v}_j = \hat{v}_j \left[ (1 - \frac{\alpha_j}{\alpha})I + \frac{\alpha_j}{\alpha} \sum_{\boldsymbol{u}_j} \hat{v}_{u_j}(\boldsymbol{u}_j) M_{X_j|\boldsymbol{u}_j} \right]$$
$$= \hat{v}_j \left[ \sum_{\boldsymbol{u}_j} \hat{v}_{u_j}(\boldsymbol{u}_j) \tilde{M}_{X_j|\boldsymbol{u}_j} \right] \quad (2)$$

To calculate $\hat{v}Q$ and project onto $x_j$, Equation 2 holds if we change $\tilde{M}_{X_j|\boldsymbol{u}_j}$ to $Q_{X_j|\boldsymbol{u}_j}$.

**Example.** *Starting distribution $p$ marginals $\hat{v}_A = \begin{bmatrix} 0.6 & 0.4 \end{bmatrix}$ and $\hat{v}_B = \begin{bmatrix} 0.5 & 0.5 \end{bmatrix}$. Multiplying this factored approximation by $M$ and then projecting can be done by multiplying $v_A$ by $\tilde{M}_A$ and $v_B$ by $0.6\tilde{M}_{B|a_0} + 0.4\tilde{M}_{B|a_1}$:*

$$\tilde{M}_A = \begin{bmatrix} .875 & .125 \\ .25 & .75 \end{bmatrix}, \tilde{M}_{B|a_0} = \begin{bmatrix} .625 & .375 \\ .5 & .5 \end{bmatrix}, \tilde{M}_{B|a_1} = \begin{bmatrix} .375 & .625 \\ .75 & .25 \end{bmatrix}.$$

## 4 Bounds for Factored Uniformization

While we have not found a suitable way of bounding the approximation error for factored RKF, we can derive bounds similar to those of the BK algorithm (Boyen and Koller, 1998) for discrete-time stochastic processes to bound the error in propagation and projection through a single $M$ matrix. However, because the process is a mixture of processes in which only a single component changes, the BK result for compound processes does not carry over.

We then use this bound to bound the error of the entire Taylor expansion and thereby the factored uniformization method.

### 4.1 Divergence Bound for Single Step

We first concentrate on bounding the error of a single multiplication by $M$. We wish to show that the KL-divergence between $v' = vM$ and $\hat{v}' = \hat{v}M$ is no greater than that between $v$ and $\hat{v}$.

We begin with a simple property of the KL-divergence. The proof is omitted, but is a consequence of the fact that entropy does not increase upon conditioning.

**Lemma 1.** *If $q(x) = \prod_i q(x_i)$ is a factored distribution and $p(x)$ is a (non-factored) distribution over the same sample space, and $x_{-i}$ is the set of all variables except $x_i$,*

$$\sum_i D_{KL}(p(x_i \mid x_{-i}) \| q(x_i)) \geq D_{KL}(p(x) \| q(x)) \ .$$

We require a mixing rate definition from Boyen and Koller (1998):

**Definition 2.** *The* mixing rate *of a stochastic matrix $M$ is defined as $\gamma \triangleq \min_{i_1, i_2} \sum_j \min(M_{i_1,j}, M_{i_2,j})$.*

We can then state that $M$ is a contraction mapping with respect to the KL-divergence:

**Theorem 3.** *Let $\gamma_i$ be the minimum (over $\mathbf{u}_i$) mixing rate of the stochastic matrix $M_{X_i | \mathbf{u}_i}$ and $\gamma = \min_i \frac{\alpha_i \gamma_i}{\alpha}$. Then,*

$$D_{KL}(v' \| \hat{v}') \leq (1 - \gamma) D_{KL}(v \| \hat{v}) \ .$$

*Proof.* If we let $v'^{(i)} = vM_i$ and $\hat{v}'^{(i)} = \hat{v}M_i$, then

$$\begin{aligned}
D_{\mathrm{KL}}(vM \| \hat{v}M) &= D_{\mathrm{KL}}(\sum_i \tfrac{\alpha_i}{\alpha} v'^{(i)} \| \sum_i \tfrac{\alpha_i}{\alpha} \hat{v}'^{(i)}) \\
&\leq \sum_i \tfrac{\alpha_i}{\alpha} D_{\mathrm{KL}}(v'^{(i)} \| \hat{v}'^{(i)}) \\
&= \sum_i \tfrac{\alpha_i}{\alpha} \big( D_{\mathrm{KL}}(v'^{(i)}_{-i} \| \hat{v}'^{(i)}_{-i}) + D_{\mathrm{KL}}(v'^{(i)}_{i|-i} \| \hat{v}'^{(i)}_{i|-i}) \big) \\
&\leq \sum_i \tfrac{\alpha_i}{\alpha} \big( D_{\mathrm{KL}}(v_{-i} \| \hat{v}_{-i}) + (1-\gamma_i) D_{\mathrm{KL}}(v_{i|-i} \| \hat{v}_{i|-i}) \big) \\
&= \sum_i \tfrac{\alpha_i}{\alpha} \big( D_{\mathrm{KL}}(v \| \hat{v}) - \gamma_i D_{\mathrm{KL}}(v_{i|-i} \| \hat{v}_{i|-i}) \big) \\
&\leq D_{\mathrm{KL}}(v \| \hat{v}) - \gamma \sum_i D_{\mathrm{KL}}(v_{i|-i} \| \hat{v}_i) \\
&\leq D_{\mathrm{KL}}(v \| \hat{v}) - \gamma D_{\mathrm{KL}}(v \| \hat{v})
\end{aligned}$$

The first inequality is from the convexity of the KL-divergence (Cover and Thomas, 1991, Theorem 2.7.2). The next inequality holds because $M_i$ does not change any variables except $x_i$ and the conditional KL-divergence of $x_i$ contracts by $\gamma_i$ (Boyen and Koller, 1998, Theorem 3). The final inequality is due to the lemma above. □

Note that unlike in BK, the global contraction rate ($\gamma$) does not depend on the in- or out-degree of model, but it is inversely proportional to $n$. This appears unfortunate, but the next section demonstrates that it is not a problem for the contraction rate of the entire Taylor expansion.

We upper-bound the increase from projection:

$$D_{\mathrm{KL}}(vM \| \hat{v}M_\perp) - D_{\mathrm{KL}}(vM \| \hat{v}M) \leq \epsilon \ .$$

As a crude upper bound, $\epsilon \leq -(n-1) \ln \eta$ where $\eta$ is the smallest marginal probability. As shown by Boyen and Koller (1999), better bounds can be placed by more careful analysis or considering the average case.

Taken together this means that after a single multiplication and projection, the KL-divergence error of our estimate can be bounded as

$$D_{\mathrm{KL}}(vM \| \hat{v}M_\perp) \leq (1 - \gamma) D_{\mathrm{KL}}(v \| \hat{v}) + \epsilon \ . \quad (3)$$

**Example.** *We have $\gamma_A = 0.5$, $\gamma_{B|a_0} = 0.5$, $\gamma_{B|a_1} = 0.17$. Therefore $\gamma = \min(0.5 \times \frac{2}{8}, 0.17 \times \frac{6}{8}) = 0.125$. Therefore the KL-divergence between the true answer and the factored approximation contracts by $1 - 0.125 = 0.825$ for each multiplication by $M$.*

### 4.2 Bound on Approximate Taylor Expansion

Our goal is not to bound the error on a single step, but rather the error of our entire approximation to the matrix exponential. Equation 3 implies that

$$D_{\mathrm{KL}}(vM^k \| \hat{v}M_\perp^k) \leq (1-\gamma)^k D_{\mathrm{KL}}(v \| \hat{v}) + \epsilon \sum_{i=0}^{k-1} (1-\gamma)^i$$

$$= (1-\gamma)^k D_{\mathrm{KL}}(v \| \hat{v}) + \epsilon \frac{1 - (1-\gamma)^k}{\gamma} \quad (4)$$

If we combine the bound from Equation 4 with the Taylor expansion of Equation 1, we can obtain a bound on the KL-divergence between the true matrix exponential, and an approximation of the Taylor expansion in which each vector of probabilities is approximated by a factored form and only the first $l$ terms are evaluated (and then renormalized):

**Theorem 4.** *Let $\alpha$, $\epsilon$, and $\gamma$ be as defined above. Let $p$ be an arbitrary distribution over the state space of the process. Let $\hat{p}$ be an arbitrary factored distribution over the same. Further, let $p' = p e^{Qt}$ be the distribution at time $t$ in the future, and let $\hat{p}' = \frac{1}{1-R_l} \sum_{k=0}^{l} e^{-\alpha t} \frac{(\alpha t)^k}{k!} \hat{p} M_\perp^k$ be the approximation of $p'$ constructed by uniformization of a Taylor expansion truncated to $l$ terms in which each matrix multiplication is projected back to the space of factored distributions. $R_l = \sum_{k=l+1}^{\infty} e^{-\alpha t} \frac{(\alpha t)^k}{k!}$. Then*

$$D_{KL}(p' \| \hat{p}') \leq e^{-\gamma \alpha t} D_{KL}(p \| \hat{p}) + \frac{\epsilon}{\gamma}(1 - e^{-\gamma \alpha t}) + R_l(\delta + \frac{\epsilon}{\gamma})$$

*where $\delta = \max_{\boldsymbol{x}} -\ln \hat{p}'_{\boldsymbol{x}}$, the maximum negative log probability over any joint assignment in the final approximate calculation.*

*Proof.* For compactness of presentation, let $\beta_k = e^{-\alpha t}\frac{(\alpha t)^k}{k!}$, $\bar{\gamma} = (1-\gamma)$, and $\bar{R}_l = (1-R_l)$. Then

$$D_{\mathrm{KL}}(p'\|\hat{p}') = D_{\mathrm{KL}}(\sum_{k=0}^{\infty}\beta_k pM^k \| \frac{1}{\bar{R}_l}\sum_{k=0}^{l}\beta_k \hat{p}M_\perp^k)$$
$$\leq \bar{R}_l D_{\mathrm{KL}}(\frac{1}{\bar{R}_l}\sum_{k=0}^{l}\beta_k pM^k \| \frac{1}{\bar{R}_l}\sum_{k=0}^{l}\beta_k \hat{p}M_\perp^k)$$
$$+ R_l D_{\mathrm{KL}}(\frac{1}{R_l}\sum_{k=l+1}^{\infty}e^{-\alpha t}\beta_k pM^k \| \frac{1}{\bar{R}_l}\sum_{k=0}^{l}\beta_k \hat{p}M_\perp^k)$$
$$\leq \sum_{k=0}^{l}\beta_k D_{\mathrm{KL}}(pM^k \| \hat{p}M_\perp^k) + R_l \delta$$
$$\leq \sum_{k=0}^{l}\beta_k [\bar{\gamma}^k D_{\mathrm{KL}}(p\|\hat{p}) + \epsilon(1-\bar{\gamma}^k)/\gamma] + R_l \delta$$
$$\leq e^{-\gamma\alpha t}D_{\mathrm{KL}}(p\|\hat{p}) + \frac{\epsilon}{\gamma}(1-e^{-\gamma\alpha t}) + R_l(\delta + \frac{\epsilon\bar{\gamma}^l}{\gamma}) .$$

The first inequality is due to the convexity of the KL-divergence in only the first argument. The second is due to the convexity of the KL-divergence in both arguments, and that the KL-divergence is bounded by the negative log of smallest probability of the second argument. The third is due to Theorem 3. The final is due to bounding a finite Taylor expansion of an exponential by the exponential. □

Note that $R_l$ goes to zero as $l$ grows. In the remaining terms, $\gamma$ almost always appears multiplied by $\alpha$. In the previous section, we commented on how $\gamma$ is inversely proportional to $n$. However, $\alpha$ is proportional to $n$ and they exactly cancel out. Thus if $l$ is large enough, we can let $\gamma' = \min_i \alpha_i \gamma_i$ and conclude

$$D_{\mathrm{KL}}(p'\|\hat{p}') \leq e^{-\gamma' t}D_{\mathrm{KL}}(p\|\hat{p}) + \frac{\alpha\epsilon}{\gamma'}(1-e^{-\gamma' t}) . \quad (5)$$

Thus the contraction rate for the full approximation is *not* a function of $n$. It is unclear how $\alpha\epsilon$ scales with $n$.

We are then left with two error terms. The first decays with $t$ and the second grows with $t$ (but is bounded). The resulting distribution $\hat{p}'$ is a mixture of factored distributions. So, if filtering is to continue past time $t$, $\hat{p}'$ must be projected back to the space of factored distributions, introducing another similar additive error. This may happen either because evidence arrives at $t$ and requires conditioning, or because we may wish to subdivide a propagation from 0 to $t$ into $m$ propagations of length $t/m$. Equation 5 implies this is not helpful for accuracy (if one considers applying the bound recursively $m$ times over intervals of length $t/m$). However, our experimental results show that some interval subdivision is helpful.

**Example.** $\gamma' = \gamma\alpha = 1$. *For large $l$, the KL-Divergence between the true distribution and the factored approximation before ($D_{KL}$) and after propagation to $t = 0.5$ ($D'_{KL}$) is related by*

$$D'_{KL} \leq 0.61 D_{KL} + (8\epsilon/1)(1-0.61)$$

*where we note that $e^{-1\times 0.5} \approx 0.61$. If the smallest possible marginal probability is $0.01$, then a crude upperbound on $\epsilon$ is $-\ln 0.01 = 4.6$.*

## 5 Adding Evidence

So far, we have discussed only how to filter without evidence. In a continuous-time process, evidence can take on two forms. First, it can be point evidence. That is, at time $t$ we observe the values of certain variables, but for no duration. We propagate to the time of the evidence, condition the distribution on the evidence, and then continue. In this case, conditioning on the evidence in expectation reduces the error (Boyen and Koller, 1998, Fact 1).

The second form of evidence is interval evidence: a variable remains in a particular state from $t_1$ to $t_2$. In this case, at $t_1$ we condition on the evidence (same as for point evidence, above). From $t_1$ until $t_2$, we monitor using a modified $Q$ matrix in which all transitions where the evidence variable changes are set to 0 (but the diagonal elements remain unchanged). The resulting $\hat{p}$ sums to the probability of the interval evidence, but it can be normalized to yield the conditional distribution of the state. The normalization makes the analysis difficult. However, we conjecture that interval evidence will also not increase the error in expectation.

Finally we note that the sparse uniformization method can have serious difficulties with evidence if none of the maintained states are consistent with the evidence.

**Example.** *Given the previous distribution $p$ for $t = 0$, and $B = b_0$ on $t = [0.5, 1)$, we propagate $p$ to $t = 0.5$ using Equation 1 (bounded), with Equation 2 to calculate the projected multiplications. We get the factored distribution $\hat{p}_A = \begin{bmatrix} 0.65 & 0.35 \end{bmatrix}$, $\hat{p}_B = \begin{bmatrix} 0.56 & 0.44 \end{bmatrix}$. We condition on $B = b_0$ by setting $\hat{p}_B = \begin{bmatrix} 1 & 0 \end{bmatrix}$. Then we similarly propagate to $t = 1$, but using*

$$Q_A = \begin{bmatrix} -1 & 1 \\ 2 & -2 \end{bmatrix}, Q_{B|a_0} = \begin{bmatrix} -3 & 0 \\ 0 & 0 \end{bmatrix}, Q_{B|a_1} = \begin{bmatrix} -5 & 0 \\ 0 & 0 \end{bmatrix}$$

*for which $\alpha_A = 2$, $\alpha_B = 5$, and $\alpha = 7$:*

$$M_A = \begin{bmatrix} 0.5 & 0.5 \\ 1 & 0 \end{bmatrix}, M_{B|a_0} = \begin{bmatrix} 0.4 & 0 \\ 0 & 1 \end{bmatrix}, M_{B|a_1} = \begin{bmatrix} 0 & 0 \\ 0 & 1 \end{bmatrix} .$$

## 6 Experimental Results

We employed two synthetic networks and a network built from a real data set in our evaluations. Since exact filtering is intractable for large models, we limited the number of variables to allow for calculations of the true approximation errors.

The synthetic networks we use are the ring and toroid dynamic Ising networks of 20 binary variables from El-Hay et al. (2010). The ring network is bidirectional, while the toroid is directed. Nodes try to track their parents with a coupling parameter, $\beta$, indicating the strength of the influence, and a rate parameter $\tau$ that is inversely proportional to the expected time between switching. We set $\tau = 4$ and $\beta = 1$. We set both networks to have a deterministic starting distribution. For the toroid the first 5 variables are in state 0 and the remaining variables are in state 1. The ring network's initial distribution is the reverse.

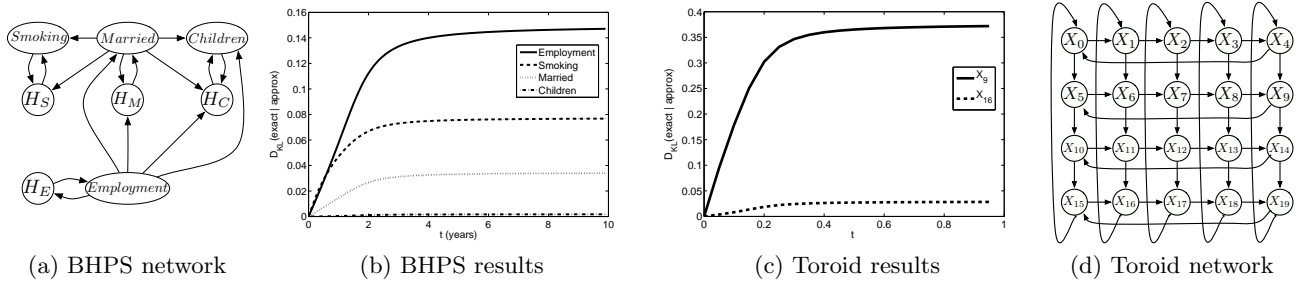

Figure 1: Accuracy versus $t$ (interval width) for the BHPS and toroid networks, for factored uniformization.

The real network we used was constructed from the British Household Panel Survey (BHPS) data set (Economic & Social Research Council, 2003). The data set records major life changes from a set of roughly 8,000 British citizens in areas including household organization, employment, income, wealth and health. We use the same network model as in Fan et al. (2010) and Nodelman et al. (2005) which chooses 4 variables: employment (student, employed, unemployed), children $(0, 1, \geq 2)$, married (not married, married) and smoking (non-smoker, smoker) and adds a hidden binary variable for each (Figure 1a). The structure and parameters of both the initial distribution and the dynamics were learned by the structural EM algorithm (Nodelman et al., 2005) and we used the learned network model for our experiments.

### 6.1 KL-Divergence Bound

Our first experiment tested the theoretic error bound. Figure 1 shows the KL-divergence between the true marginals and the marginals computed by factored uniformization for the BHPS network and the toroid network. The bound on the KL-divergence of the full distribution is also a bound on any marginal, and experimentally the errors on the marginals grow initially and then asymptote, as per Theorem 4.

### 6.2 Approximation Comparison

We then compared our factored uniformization method ($U_F$) to other approximations. In particular, we compared to factored RKF ($RKF_F$), as explained in Section 2.1, sparse uniformization ($U_S$), and the mean field (MF) approach of Cohn et al. (2009) for CTBNs. For the purpose of comparison, the MF method was extended to accept evidence on subsets of variables.

We varied an error tolerance parameter for each method to map the trade-off between error and runtime. For $U_F$, we varied the uniformization-specific parameter $\theta$ that determines the number of intervals of propagation (see Sidje et al., 2007). We fixed the number of terms of the Taylor series expansion ($l$) to a value that performed reasonably. Similarly, for sparse uniformization we varied $\theta$ and fixed $l$ to a well-performing value. MF and $RKF_F$ both use RKF for integration, so we varied the error tolerance of RKF.

The evidence for the ring and toroid networks was set to be relatively unexpected: for $t \in [0.5, 1.0)$ $x_0 = 1$ and $x_1 = 0$. For the ring, we queried the distribution of $x_{10}$ (far from the evidence) and $x_{19}$ (adjacent to the evidence) at time $t = 1$. For the toroid, we queried the distribution of $x_6$ (adjacent to the evidence) and $x_{13}$ (a node more the in middle) at $t = 1$. For the BHPS network, we chose evidence where employment is observed to be in the student state continuously from year 1 to year 5. We queried the marginal distribution of the variables smoking and children at $t = 5$ years.

Figure 2 shows the KL-divergence between the true and approximated marginals of the query variable for each of the three networks for each query. These results are typical of other query marginals. In general, factored uniformization ($U_F$) performs the best, especially for smaller running times (with $U_S$ occasionally being the method of choice for longer running times). However, when the query node is close to the evidence, $RKF_F$ performs better ($x_{10}$ for the ring and $x_{13}$ for the toroid network are examples). The advantage of $U_S$ for longer running times is predictable as it approaches the true value as more states are retained. Finally, note that for the BHPS query over the children variable, $U_S$ has infinite KL-divergence (and hence does not appear on the plot). This is because all of the retained states have this variable equal to 0. While this is certainly the most common value at $t = 5$, it does not have probability 1 under the true distribution and thus the KL-divergence between the true distribution and the approximation of $U_S$ is infinite.

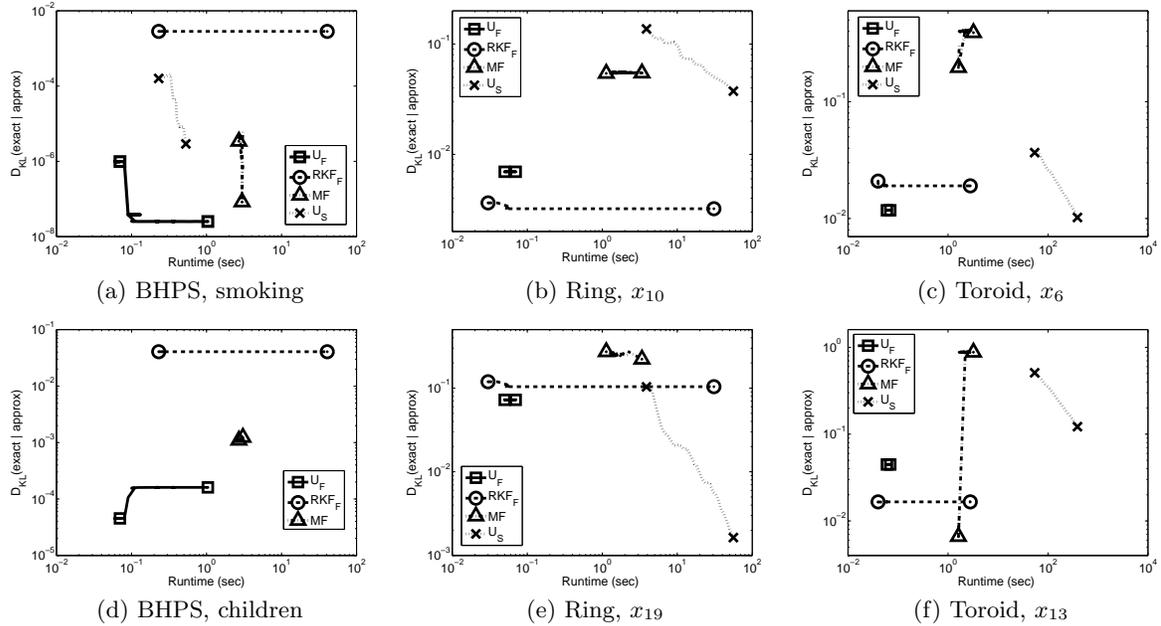

Figure 2: Computation time versus accuracy comparisons.

## 7 Conclusion

We have demonstrated approximate continuous-time filtering based on uniformization. It is simple to implement, and we have proven bounds on the KL-divergence of its error. The approximation can be made more accurate by lumping together variables into joint marginals. The bounds are similar in style to those of BK and also depend on the mixing time of the individual components (adjusted for the continuous-time nature of the system). Our experimental results demonstrate that the theoretic bounded error holds in practice. Furthermore, our method gives superior time-accuracy trade-offs for most of the examples tested in this paper.